\documentclass[nonacm]{acmart}

\setcopyright{none}
\copyrightyear{}  
\acmYear{2026}
\acmDOI{}
\acmISBN{}

\usepackage{tabularx}
\usepackage{booktabs}
\usepackage{cleveref}
\usepackage{enumerate}
\usepackage{listings}
\usepackage{multirow}
\usepackage{siunitx}
\usepackage[textsize=tiny]{todonotes}
\usepackage[normalem]{ulem}
\usepackage{xspace}
\usepackage{soul}
\usepackage{tikz}
\usepackage{subcaption}
\usepackage[table]{xcolor}
\usetikzlibrary{arrows.meta,positioning,shapes.geometric}

\newcommand{\dataDockerImages}{\num{3620}\xspace}

\lstdefinelanguage{docker}{
  keywords={FROM, RUN, COPY, ADD, ENTRYPOINT, CMD,  ENV, ARG, WORKDIR, EXPOSE, LABEL, USER, VOLUME, STOPSIGNAL, ONBUILD, MAINTAINER, HEALTHCHECK},
  keywordstyle=\color{blue}\bfseries,
  identifierstyle=\color{black},
  sensitive=false,
  comment=[l]{\#},
  commentstyle=\color{purple}\ttfamily,
  stringstyle=\color{red}\ttfamily,
  morestring=[b]',
  morestring=[b]"
}

\lstdefinelanguage{ghactions}{
  keywords={steps, uses, with, toolchain, override, run, name, path},
  keywordstyle=\color{blue}\bfseries,
  identifierstyle=\color{black},
  sensitive=false,
  comment=[l]{\#},
  commentstyle=\color{purple}\ttfamily,
  stringstyle=\color{red}\ttfamily,
  morestring=[b]',
  morestring=[b]"
}

\begin{document}

\newcommand\SelectedRuns{6566} 
\newcommand\SuccessfulRuns{5298\xspace} 
\newcommand\TotalRepos{705} 
\newcommand\NumberOfImagesBuilding{3836} 
\newcommand\PropOfImagesBuilding{72} 
\newcommand\ImagesBitwiseReproducible{4} 
\newcommand\TotalImagesBitwiseComp{1096} 
\newcommand\AverageBitwiseComp{40.3} 
\newcommand\MedianBitwiseComp{34.4} 
\newcommand\ImagesFuncExactReproducible{70} 
\newcommand\ImagesFuncExactNonReproducible{1026} 
\newcommand\ImagesFuncExactReproducibleProp{6} 
\newcommand\TotalImagesFuncComp{1096} 
\newcommand\AverageFuncExactComp{27.6} 
\newcommand\MedianFuncExactComp{22.6} 
\newcommand\ImagesFuncMinorReproducible{271} 
\newcommand\ImagesFuncMinorReproducibleProp{25} 
\newcommand\AverageFuncMinorComp{12.9} 
\newcommand\MedianFuncMinorComp{3} 
\newcommand\ImagesFuncMajorReproducible{406} 
\newcommand\ImagesFuncMajorReproducibleProp{37} 
\newcommand\AverageFuncMajorComp{7} 
\newcommand\MedianFuncMajorComp{1} 
\newcommand\RRebuildability{0.05} 
\newcommand\RBitwise{0.27} 
\newcommand\ImagesFuncPackagesReproducible{573} 
\newcommand\TrivyNotAuditable{20} 
\newcommand\TrivyTotal{1116} 
\newcommand\TrivyWorks{1096} 
\newcommand\RFuncExact{0.12} 
\newcommand\RFuncMinor{0.10} 
\newcommand\RFuncMajor{0.08} 

\title{Docker Does Not Guarantee Reproducibility}

\author{Julien Malka}
\email{julien.malka@telecom-paris.fr}
\orcid{0009-0008-9845-6300}
\affiliation{\institution{LTCI, Télécom Paris, Institut Polytechnique de Paris}
 \city{Palaiseau}
 \country{France}
}
\author{Stefano Zacchiroli}
\email{stefano.zacchiroli@telecom-paris.fr}
\orcid{0000-0002-4576-136X}
\affiliation{\institution{LTCI, Télécom Paris, Institut Polytechnique de Paris}
 \city{Palaiseau}
 \country{France}
}
\author{Théo Zimmermann}
\email{theo.zimmermann@telecom-paris.fr}
\orcid{0000-0002-3580-8806}
\affiliation{\institution{LTCI, Télécom Paris, Institut Polytechnique de Paris}
 \city{Palaiseau}
 \country{France}
}

\begin{abstract}
The reproducibility of software environments is a critical concern in modern software engineering, with ramifications ranging from the effectiveness of collaboration workflows to software supply chain security and scientific reproducibility.
Containerization technologies like Docker address this problem by encapsulating software environments into shareable filesystem snapshots known as images.
While Docker is frequently cited in the literature as a tool that enables reproducibility in theory, the extent of its guarantees and limitations in practice remains under-explored.

In this work, we address this gap through two complementary approaches. First, we conduct a systematic literature review to examine how Docker is framed in scientific discourse on reproducibility and to identify documented best practices for writing Dockerfiles enabling reproducible image building.
Then, we perform a large-scale empirical study of \SuccessfulRuns Docker builds collected from GitHub workflows. 
By rebuilding these images and comparing the results with their historical counterparts, we assess the real reproducibility of Docker images and evaluate the effectiveness of the best practices identified in the literature.
\end{abstract}


\maketitle
\section{Introduction}


The \emph{reproducibility of build environments}~\cite{courtes-repro-env-2015, malka-repro-space-time-2024}---i.e., the ability to recreate the technical environment needed to develop and deploy a specific piece of software---is a fundamental aspect of modern software engineering, with significant implications for both developers and users.
For developers, build environment reproducibility enables reliable recompilation of software from source code---even years after its initial creation---facilitating collaboration, long-term maintenance, and debugging.
For end users, the ability to rebuild software from source can provide 
security guarantees, as it avoids the need of trusting opaque binaries built by third parties~\cite{lamb_reproducible_2022}.

Yet, the growing complexity of software components and  their dependencies~\cite{decan_empirical_2019}, driven by composability and modular design, makes reproducing exact build environments increasingly challenging.
This issue, which contributes to the broader \emph{reproducibility crisis}~\cite{hassan_characterising_2025, perkel_challenge_2020} issue, is well-documented in scientific computing~\cite{ivie-repro-sci-comp-2018, stodden-repro-sci-computing-2012}, where re-running old code to validate or extend prior work is an integral part of the methodology.
Yet the problem extends beyond academia and touches all fields of software engineering.

Containerization technologies~\cite{bentaleb-container-taxo-2022} allow software environments to be encapsulated and packaged as filesystem snapshots known as \emph{container images}, usually in the OCI standard format~\cite{noauthor_open_nodate}, which can then be shared and executed on other systems.
Docker~\cite{docker-web} is a widely adopted open-source container runtime and build tool that popularized the use of build recipes called \emph{Dockerfiles} to create container images.
The straightforward syntax and developer-friendly tooling of Docker contributed to its wide adoption across the open source ecosystem, to the point that a 2020 study mining Dockerfiles from the World of Code archive found that over 1.9 million distinct repositories used them~\cite{eng_revisiting_2021}.
Docker supports a broad range of workflows in software engineering---including building, packaging, distributing, and deploying software---by making software environments transportable across systems.
Thanks to this portability aspect, Docker is frequently recommended in both industry and academia for improving the reproducibility of computational environments~\cite{nust_ten_2020}.

However, Docker images are build artifacts, which suffer from common limitations for that type of software artifacts:
\begin{itemize}
\item they are large, binary artifacts that must be explicitly stored and archived, to avoid becoming unavailable over time~\cite{peikert_reproducible_2021};
\item they are typically tied to specific hardware architectures, limiting their portability~\cite{boettiger_introduction_2015};
\item they contain arbitrary binary artifacts that are difficult to audit and can represent a security risk~\cite{osorio_dockerpedia_2022,morris_use_2017};
\item modifying them to upgrade contained artifacts or change their behavior is impractical~\cite{osorio_dockerpedia_2022}.
\end{itemize}

To address these limitations, it is essential to be able to \emph{reliably rebuild a Docker image from its  Dockerfile}, ensuring that the rebuilt image matches the original one.
Different meanings of ``matches'' can be considered for this need.

When images can be rebuilt in a way that produces \emph{bitwise-identical} copies (as in the \emph{reproducible builds} approach~\cite{lamb_reproducible_2022}), trusting third-party distributors like public image registries become easier, strengthening the security of the software supply chain.
Even when bitwise reproduction is not achievable, \emph{functional equivalence}---where the rebuilt image contains the same software components, at the same versions, of the original image---remains an essential requirement.
Failing it, discrepancies in image content may lead to observable differences when using the image.

\paragraph*{Research questions}
In this work, we study Docker as a tool to enable the reproducibility of software environments, with a particular focus on the \emph{gap between theory and practice}: what is Docker \emph{believed} to do in that respect versus what it \emph{actually} does.
Specifically, we address the following research questions:
\begin{enumerate}[\bfseries RQ1:]
\item What are the claims and recommendations in the scientific literature on the reproducibility of Docker images?
\item To what extent can Docker images built in the past be reproduced today from their Dockerfiles, to various degrees of equivalence (bitwise identical, functionally equivalent)?
\item Are the recommendations present in the scientific literature about Docker image reproducibility effective?
\end{enumerate}

\paragraph*{Contributions}
We make two primary contributions:
\begin{enumerate}[(1)]
\item We perform a \textbf{systematic literature review (SLR)} of scientific literature to identify how Docker is discussed in the context of reproducibility and map documented best practices for writing Dockerfiles that, allegedly, enable image reproducibility (RQ1).
\item We conduct \textbf{a large-scale empirical study of Docker image reproducibility}.
  Leveraging a dataset of GitHub workflow runs, we reproduce and analyze~\SuccessfulRuns historical Docker builds captured in October 2023.
  By comparing rebuilt images with their historic counterparts, we quantify the level of reproducibility of real-world Docker builds (RQ2).
  Then we empirically assess how the best practices for Docker reproducibility documented in the literature are used by developers, and whether they lead to improved reproducibility (RQ3).
\end{enumerate}

\textbf{Data availability statement:} We make our code required to run our analysis available in the replication package~\cite{this-replication-package}, which we permanently archive on Software Heritage.

\section{Background}
\label{sec:context}

\subsection{Software reproducibility}

This article is concerned with the notion of \emph{reproducibility} in software engineering. Reproducibility can take various meanings, and in this work, we explore several of them.

The strictest notion of reproducibility is \emph{bitwise reproducibility}, a.k.a. \emph{reproducible builds}. It is defined as the ability to reproduce the exact same binary artifact (bitwise identical) from a fixed source code and build environment. This definition has the advantage of being easy to verify, and it is thus useful both from a software supply chain security perspective~\cite{lamb_reproducible_2022}, and from a scientific perspective~\cite{vila_impact_2024}, to show that the source code accompanying a scientific publication can be used to reproduce the results of the publication.

From a software engineering perspective, however, more relaxed definitions of reproducibility are just as useful. For instance, \emph{functional reproducibility} focuses on the behavior of the software rather than its exact binary representation. Therefore, when comparing two environments, we will consider them functionally equivalent if they contain the same software components, at the same versions. What to consider as ``same versions'' can then be discussed. If we want bug-by-bug reproducibility, we should use strictly the same versions. But if we are only interested in compatibility, we can relax this constraint to same minor version, or even same major version.

Finally, rebuildability is an even looser notion of reproducibility, requiring only that the build process still runs successfully. In Docker’s context, this means we can still build an image from the Dockerfile, even if the result is not bitwise identical or functionally equivalent to the original. This remains useful for reuse, though it offers fewer guarantees about the relationship between the resulting software environment and the historical one.

\subsection{Containers, Docker images and Dockerfiles}

A \emph{container} is a lightweight, isolated execution environment that runs a process with its own filesystem, networking, and process namespace, while sharing the host operating system kernel~\cite{bentaleb-container-taxo-2022}.
Containers are designed for fast startup, resource efficiency, and high portability.
A Docker~\cite{docker-web} image is a standardized, self-contained package that includes all the components needed to run an application in a container: executable binaries, system libraries, configuration files, etc.
Internally, it is an immutable filesystem snapshot made of a superposition of read-only \emph{layers}. Each layer encodes a set of filesystem changes such as file additions, modifications, or deletions. This layering model enables efficient storage and distribution: shared layers can be cached and reused across images.
Container images follow the Open Container Initiative (OCI) image specification~\cite{noauthor_open_nodate} and may be produced by various build tools, but the most widely adopted one is the built-in Docker builder.

The Docker builder takes as input a build recipe called a \emph{Dockerfile}.
Dockerfiles specify how to construct an image from a sequence of instructions, such as selecting a base image, copying files, installing packages, and defining environment variables. For example:
\begin{lstlisting}[caption=Example of a Dockerfile describing steps to build a container for a Python application.,captionpos=b,basicstyle=\small\ttfamily,language=docker]
FROM python:3.13-slim
COPY requirements.txt .
RUN pip install -r requirements.txt
COPY . /app
CMD ["python", "/app/main.py"]
\end{lstlisting}
This Dockerfile creates an image from a minimal Python base, installs project dependencies, copies the application code, and defines the startup command.
Dockerfiles are imperative and order-sensitive: each instruction creates a new image layer that may affect later build steps.
Each layer is content-addressed by a cryptographic hash of its contents and metadata. To improve reproducibility, the Docker builder supports the \texttt{SOURCE\_DATE\_EPOCH} variable, which standardizes timestamps in generated files and metadata, helping produce identical hashes across builds.

\subsection{GitHub actions}

GitHub Actions~\cite{chandrasekara_2021_ghactions} is a continuous integration and deployment (CI/CD) service natively integrated with the GitHub collaborative development platform. It enables developers to automate workflows triggered by repository events---such as code pushes or pull requests---by executing a series of commands in virtual machines (called \emph{runners}) provided by GitHub.
Workflows are specified using YAML files stored directly in the repository. Each workflow consists of one or more \emph{jobs}, in turn composed of ordered \emph{steps}. Steps run sequentially within the same job environment, allowing state propagation---e.g., via the filesystem or environment variables---making step chaining key for composing logic in CI pipelines.
Each step may invoke a shell command or call a reusable, modular unit of functionality known as a \emph{GitHub Action}.
Actions are typically provided by other GitHub repositories, are versioned, and configurable.
They encapsulate common tasks (such as checking out code, setting up a programming language environment, or uploading artifacts), promoting reuse 
across repositories and workflows.

An example of a GitHub Action is \texttt{docker/build-push-action}, a widely used action maintained by Docker and designed to build container images from Dockerfiles and optionally push them to a registry.
It supports numerous parameters to customize the Docker build, which are translated to flags passed to the \texttt{docker build} command.
Here is an excerpt of a workflow using this action:
\\
\begin{lstlisting}[caption=Example of a simple workflow building a Docker image from a GitHub repository.,captionpos=b,basicstyle=\small\ttfamily,language=ghactions]
steps:
  - uses: actions/checkout@v4
  - uses: docker/build-push-action@v6
    with:
      context: .
      push: true
      tags: myname/myapp:latest
\end{lstlisting}

In this example, the official GitHub checkout action is used to clone the repository, then the \texttt{docker/build-push-action} builds a Docker image from a Dockerfile at the root of the repository. This workflow could be further simplified because the Docker action supports doing the repository checkout by itself.

\section{Systematic literature review}
\label{sec:slr}

To answer RQ1, we conducted a systematic literature review (SLR), adhering to established best practices as outlined in the ACM SIGSOFT \emph{Empirical Standards for Software Engineering}~\cite{ralph_empirical_2021}.
This SLR is a scoping and critical review~\cite{ralph_paving_2022}, whose primary objective is to extract and categorize the academic discourse on Docker reproducibility, with a focus on identifying expressed beliefs, and recommendations for writing Dockerfiles that enable reproducible image building.
SLR outcomes will serve as foundations for an empirical study, where we evaluate to which extent these beliefs and recommendations are founded (RQ2, RQ3).
We start by refining RQ1 into the following sub-questions:
\begin{enumerate}
\item[\textbf{RQ1.a:}] What are the expectations, implicit or explicit, present in the scientific literature on the reproducibility guarantees provided by Docker?
\item[\textbf{RQ1.b:}] What are the common causes, documented in the scientific literature, of non-reproducibility of Docker images?
\item[\textbf{RQ1.c:}] What are the recommendations, put forward in the 
literature, to maximize the reproducibility of Docker images?
\end{enumerate}

\subsection{Methodology}

\subsubsection{Search protocol}

Before starting the systematic search, we performed an informal literature review to identify key papers and common terminology used in the scientific discourse on Docker reproducibility. Based on this preliminary work, we identified several domains where Docker reproducibility is discussed, including in particular scientific computing and software engineering, and some common keywords used in these domains.

Next, we went through a prototyping phase to define our search strategy.
We performed initial queries across standard scientific databases including Scopus, IEEE Xplore, the ACM Digital Library, and Springer.
However, because the search is limited to title and abstract on several of these databases, our queries returned very few results.
This suggests that only few publications address Docker reproducibility as their main contribution, and that a full content search was necessary to identify relevant discussions occurring in broader contexts.
For that purpose, we switched to Google Scholar, which supports full-text search queries.
This also allowed us to find more types of scientific literature such as tutorials, theses, and preprints, which is recommended in the SIGSOFT Empirical Standards for Systematic Literature Reviews~\cite{ralph_empirical_2021}. To select our search keywords, we aimed to cover the scientific fields of provenance that we had previously identified in the informal literature review. We used four search queries, detailed in \Cref{tab:search_queries}.

\begin{table}[h]
\caption{Search queries used in Google Scholar for the systematic literature review.}
\label{tab:search_queries}
\centering
\small
\begin{tabular}{p{5cm}|c|p{6cm}}
\toprule
\textbf{Keywords} & \textbf{Results} & \textbf{Justification} \\
\midrule
``reproducible science'' + dockerfile & 149 & Targets the field of scientific computing \\
``reproducible build environment'' + docker & 13 & Targets the field of software engineering, with a focus on reproducible environments \\
reproducibility + smells + dockerfile & 150 & Targets the field of software engineering, with a focus on Dockerfile quality \\
reproducibility + dockerfile + linter & 220 & Targets the field of software engineering, with a focus on Dockerfile quality \\
\bottomrule
\end{tabular}
\end{table}

While the first search query is aimed at finding relevant works from the scientific computing literature, the third and the last are using specific keywords that we know to be common in the software engineering literature. We checked that all the relevant works we knew about were covered in the search results, and that the number of papers found in the first phase was reasonable.

All searches were performed in May 2025. We collected the results of each query in a separate folder of a Zotero collection, which we make publicly available to facilitate inspection and replication~\cite{this-slr-zotero}.

\subsubsection{Inclusion and exclusion criteria}

Following the search phase, we removed 82 exact duplicates that appeared in the results of several search queries (by merging the entries in Zotero), leaving 450 unique articles.
We then applied a three-stage filtering process based on the title, abstract, and full-text content.
At each phase, the first and last authors devised the inclusion and exclusion criteria by processing a subset of entries together, discussing potential criteria until each entry was included or excluded without ambiguity.
The criteria were intentionally designed to exclude articles that were clearly irrelevant, while retaining those with uncertainty for further evaluation in subsequent filtering phases.
Once criteria were stabilized, the first and last authors shared the workload of checking individual articles.

\begin{table}
  \caption{Filtering criteria used to select articles for the systematic
    literature review on Docker reproducibility.}
  \label{tab:filtering_criteria}
  \small
  \begin{tabular}{l|l|l}
    \toprule
    \textbf{Stage} & \textbf{Type} & \textbf{Criterion} \\
    \midrule

    \multirow{4}{*}{Title} & \multirow{3}{*}{Inclusion} & \textbf{IT1:} Title contains keywords related to reproducibility \\
                   & & \textbf{IT2:} Title contains keywords related to Docker or containers \\
                   & & \textbf{IT3:} Title contains keywords related to
                       infrastructure as code \\
    \cline{2-3}
    & \multirow{1}{*}{Exclusion} & \textbf{EC1:} Title does not contain any keyword mentioned in inclusion criteria \\

    \midrule
    \multirow{3}{*}{Abstract} & \multirow{1}{*}{Inclusion} & \textbf{IA1:} Contribution explicitly related to Docker or reproducibility best practices \\
    \cline{2-3}
    & \multirow{3}{*}{Exclusion} & \textbf{EA1:} Contribution not related to Docker \\
                   && \textbf{EA2:} Only uses Docker without reflection on Docker as a tool \\

    \midrule
    \multirow{6}{*}{Full text} & \multirow{3}{*}{Inclusion} & \textbf{IC1:} Contains recommendations on Dockerfile best practices \\
                   & & \textbf{IC2:} Recommends or advises against using Docker for reproducibility \\
                   & & \textbf{IC3:} Addresses smells or bad practices in Dockerfiles \\
    \cline{2-3}
    & \multirow{3}{*}{Exclusion} & \textbf{EC1:} Article full text not accessible to authors \\
                   & & \textbf{EC2:} No references or only anecdotal ones to Dockerfiles \\
                   & & \textbf{EC3:} Tool contribution without reproducibility or quality focus \\

    \bottomrule
  \end{tabular}
\end{table}

\Cref{tab:filtering_criteria} summarizes the criteria applied at each stage.
During title screening, we retained articles whose titles mentioned reproducibility, Docker or containers, or infrastructure as code.
This step eliminated 271 articles. 
Next, we reviewed the abstracts of the remaining articles.
We retained only those that explicitly discussed Docker as part of their contributions or addressed best practices related to reproducibility in software engineering or scientific workflows. Articles that merely used Docker (e.g., to provide a replication package) without reflecting on its role as a tool were excluded. After this phase, 83 articles remained. Finally, during the full-text screening phase, articles were included if they offered actionable recommendations for writing Dockerfiles, discussed the use of Docker for reproducibility, or discussed Dockerfile smells and anti-patterns. We excluded articles that were inaccessible, lacked any relevant mention of Dockerfiles, or presented Dockerfile generating tools without any focus on reproducibility or quality aspects.
We additionally excluded a single Japanese-language master thesis because we don't understand the language and did not want to rely on automated translation, but the relevant contribution from the master thesis was also presented in an English-language paper that we did include. The final set of articles is mainly in English, with the exception of two works (SC7 and O3) written in French.
The Zotero collection contains tags indicating the filtering decision for each article at each phase, and the inclusion/exclusion criteria that were applied.

\subsubsection{Backward snowballing}

In order to mitigate the risk that our search queries would not cover all the space of papers relevant to answer our research questions, we performed backward snowballing by examining the related work section (or the introduction, when there was no related work section, or the full text for literature reviews) of all papers selected after the abstract filtering phase. References cited in these sections were considered for snowballing based on their title and their description in the paper that cites them. This process yielded 26 additional papers, 13 of which met our inclusion criteria. Additionally, we included the official Docker documentation on best practices for building containers, as it was frequently cited in the selected literature.
We ended up with a total of 60 papers to analyze.

\begin{table}
  \caption{Scientific Computing papers included in the SLR}
  \label{tab:sc_papers}
  \small
  \begin{tabular}{l|p{6cm}|p{2cm}|l|p{2.4cm}|l}
    	\textbf{Code} & \textbf{Title} & \textbf{Authors} & \textbf{Year} & \textbf{Domain} & \textbf{Category} \\
    \midrule
    SC1 & A reproducible data analysis workflow with R Markdown, Git, Make, and Docker & Peikert et al. & 2021 & Psychology & Journal \\
    SC2 & A Reproducible Tutorial on Reproducibility in Database Systems Research & Fischer et al. & 2024 & Database Systems & Conference \\
    SC3 & An introduction to Docker for reproducible research & Boettiger & 2015 & General & Journal \\
    SC4 & Computational Reproducibility via Containers in Psychology & Clyburne-Sherin et al. & 2019 & Psychology & Journal \\
    SC5 & Bringing your tools to CyVerse discovery environment using Docker & Devisetty et al. & 2016 & Life Sciences & Journal \\
    SC6 & Burner: Recipe Automatic Generation for HPC Container Based on Domain Knowledge Graph & Zhong et al. & 2022 & High Performance Computing & Journal \\
    SC7 & Comment rater la reproductibilité de ses expériences? & Guilloteau et al. & 2023 & General & Conference \\
    SC8 & Containers for computational reproducibility & Moreau et al. & 2023 & General & Journal \\
    SC9 & containerit: Generating Dockerfiles for reproducible research with R & Nüst et al. & 2019 & General & Journal \\
    SC10 & Introduction to Containers \& Application to AI at LRZ & Dufour et al. & 2023 & Artificial Intelligence & Textbook \\
    SC11 & Using Docker Containers to Improve Reproducibility in Software and Web Engineering Research & Cito et al. & 2016 & Web Engineering & Conference \\
    SC12 & Leveraging Containers for Reproducible Psychological Research & Wiebels et al. & 2021 & Psychology & Journal \\
    SC13 & Recommendations for the packaging and containerization of bioinformatics software & Bjorn et al. & 2019 & Bioinformatics & Journal \\
    SC14 & Reproducible image handling and analysis & Miura et al. & 2021 & Life Sciences & Journal \\
    SC15 & The Dockstore: enabling modular, community-focused sharing of Docker-based genomics tools and workflows & O'Connor et al. & 2017 & Bioinformatics & Journal \\
    SC16 & Ten simple rules for writing Dockerfiles for reproducible data science & Nüst et al. & 2020 & Bioinformatics & Journal \\
    SC17 & Toward practical transparent verifiable and long-term reproducible research using Guix & Vallet et al. & 2022 & General & Journal \\
    SC18 & Use of Docker for deployment and testing of astronomy software & Morris et al. & 2017 & Astronomy & Journal \\
    \bottomrule
  \end{tabular}
\end{table}

\begin{table}
  \caption{Software Engineering papers included in the SLR}
  \label{tab:se_papers}
  \small
  \begin{tabular}{l|p{8cm}|p{2cm}|l|p{2.1cm}}
    	\textbf{Code} & \textbf{Title} & \textbf{Authors} & \textbf{Year} & \textbf{Venue} \\
    \midrule
    SE1 & A large-scale data set and an empirical study of Docker images hosted on Docker Hub & Lin et al. & 2020 & ICSME \\
    SE2 & An empirical case study on the temporary file smell in dockerfiles & Lu et al. & 2019 & IEEE Access \\
    SE3 & A multi-dimensional analysis of technical lag in Debian-based Docker images & Zerouali et al. & 2021 & Empir. Softw. Eng. \\
    SE4 & An empirical analysis of the docker container ecosystem on github & Cito et al. & 2017 & MSR \\
    SE5 & Recommending Base Image for Docker Containers based on Deep Configuration Comprehension & Zhang et al. & 2022 & SANER \\
    SE6 & An empirical study on self-admitted technical debt in Dockerfiles & Azuma et al. & 2022 & Empir. Softw. Eng. \\
    SE7 & One size does not fit all: an empirical study of containerized continuous deployment workflows & Zhang et al. & 2018 & ESEC/FSE \\
    SE8 & An insight into the impact of dockerfile evolutionary trajectories on quality and latency & Zhang et al. & 2018 & ICCSA \\
    SE9 & Assessing and Improving the Quality of Docker Artifacts & Rosa et al. & 2022 & ICSME \\
    SE10 & Characterizing the occurrence of dockerfile smells in open-source software: An empirical study & Wu et al. & 2020 & IEEE Access \\
    SE11 & Structured information on state and evolution of dockerfiles on github & Schermann et al. & 2018 & MSR \\
    SE12 & Dockercleaner: Automatic repair of security smells in dockerfiles & Bui et al. & 2023 & ICSME \\
    SE13 & Dockerfile Flakiness: Characterization and Repair & ShabaniMirzaei & 2024 & Master Thesis \\
    SE14 & Implementing a container-based build environment: a case study & Pärssinen & 2020 & Master Thesis \\
    SE15 & DockerPedia: A Knowledge Graph of Software Images and Their Metadata & Osorio et al. & 2022 & Intl. J. Soft. Eng. \& Knowledge Eng. \\
    SE16 & Doctor: Optimizing Container Rebuild Efficiency by Instruction Reordering & Zhu et al. & 2025 & ISSTA \\
    SE17 & DRIVE: Dockerfile Rule Mining and Violation Detection & Zhou et al. & 2023 & TOSEM \\
    SE18 & Empirical Study of the Docker Smells Impact on the Image Size & Durieux & 2024 & ICSE \\
    SE19 & Fixing Dockerfile smells: an empirical study & Rosa et al. & 2024 & Empir. Softw. Eng. \\
    SE20 & Improving the Developer Experience of Dockerfiles & da Silva Matos & 2023 & Master Thesis \\
    SE21 & Latest Image Recommendation Method for Automatic Base Image Update in Dockerfile & Kitajima et al. & 2020 & ICSOC \\
    SE22 & Learning from, understanding, and supporting DevOps artifacts for docker & Henkel et al. & 2020 & ICSE \\
    SE23 & Meta-Maintanance for Dockerfiles: Are We There Yet? & Tanaka et al. & 2023 & Preprint \\
    SE24 & Not all Dockerfile Smells are the Same: An Empirical Evaluation of Hadolint Writing Practices by Experts & Rosa et al. & 2024 & MSR \\
    SE25 & Outdated software in container images & Linnalampi & 2021 & Master Thesis \\
    SE26 & Revisiting Dockerfiles in Open Source Software Over Time & Eng et al. & 2021 & MSR \\
    SE27 & Refactoring for Dockerfile Quality: A Dive into Developer Practices and Automation Potential & Ksontini et al. & 2025 & Preprint \\
    SE28 & RUDSEA: recommending updates of Dockerfiles via software environment analysis & Hassan et al. & 2018 & ASE \\
    SE29 & The Docker Hub image inheritance network: Construction and empirical insights & Opdebeeck et al. & 2023 & SCAM \\
    SE30 & Shipwright: A human-in-the-loop system for dockerfile repair & Henkel et al. & 2021 & ICSE \\
    SE31 & Understanding and Predicting Docker Build Duration: An Empirical Study of Containerized Workflow of OSS Projects & Wu et al. & 2022 & ASE \\
    \bottomrule
  \end{tabular}
\end{table}

\begin{table}
  \caption{Other sources included in the SLR}
  \label{tab:o_papers}
  \small
  \begin{tabular}{l|p{6.7cm}|p{2.2cm}|l|l}
    	\textbf{Code} & \textbf{Title} & \textbf{Authors} & \textbf{Year} & \textbf{Category / Venue} \\
    \midrule
    O1 & Best practices & Docker Docs & 2025 & Webpage \\
    O2 & Cloud security with Docker and Kubernetes & Raftopoulos & 2025 & Master Thesis \\
    O3 & Déploiements reproductibles dans le temps avec GNU Guix & Courtès & 2021 & GNU/Linux Magazine \\
    O4 & Representing Dockerfiles in RDF & Tommasini et al. & 2017 & Conference (ISWC) \\
    O5 & Docker and Kubernetes for Java Developers & Krochmalski & 2017 & Practitioner Book \\
    O6 & From containers to Kubernetes with Node. js & Juell & 2020 & Practitioner Book \\
    O7 & Infrastructure as Code (IAC) Cookbook & Jourdan et al. & 2017 & Practitioner Book \\
    \bottomrule
  \end{tabular}
\end{table}

\subsubsection{Content analysis}

We adopted a qualitative content analysis approach inspired by Charmaz constructive grounded theory~\cite{charmaz_constructing_2014}, structured in two main phases. In the \emph{initial coding phase}, the first and last authors shared the workload of reading the full texts of the selected articles and identified excerpts relevant to the research questions. During this open-ended process, excerpts were assigned short, descriptive codes that captured key ideas, practices, or concerns related to Docker and reproducibility. During this phase, when multiple versions (non-exact duplicates) of a paper existed (4 cases with two versions), they reviewed one in depth, then screened the other to check for any additional relevant content.
In the \emph{focused coding phase}, all authors examined together the obtained codes in a card-sorting session to uncover patterns and similarities across the dataset. They discussed and agreed on a new set of refined codes to form a structured taxonomy of academic recommendations and beliefs related to Dockerfile quality and the reproducibility of containerized software environments.
Note that we do not report inter-rater reliability metrics, as our critical literature review follows a constructivist, rather than a positivist, paradigm and the coding does not involve independent coding by multiple researchers, but rather a collaborative and iterative refinement of codes.

In this phase, we also sorted the papers into two main classes: software engineering papers, where Docker is the \emph{object of study}, and scientific computing papers, where Docker is used as a \emph{tool} to provide reproducibility for scientific workflows. We obtained (after removal of the 4 non-exact duplicates) 31 software engineering papers (denoted SE1 to SE31, and listed in \Cref{tab:se_papers}) and 18 scientific computing papers (denoted SC1 to SC18, and listed in \Cref{tab:sc_papers}). The other papers are denoted O1 (the official Docker recommendations) to O7 (and listed in \Cref{tab:o_papers}).

In our replication package~\cite{this-replication-package}, we provide a PDF document with all the excerpts from the initial coding phase, presented in context. This is the document which was printed and cut in pieces for the collaborative card-sorting / focused coding session. We also provide a YAML file containing the complete taxonomy of codes obtained in the focused coding phase, along with the mapping of codes to relevant quotes from the papers.

\subsection{Results}

Our results in this section are mainly qualitative, but we start by providing some quantitative statistics about the papers included in the SLR.
The 31 software engineering papers (SE) include 7 journal articles, 18 conference papers, 4 master theses, and 2 preprints.
The most represented venues are Mining Software Repositories (MSR) with 4 papers, followed by the International Conference on Software Engineering (ICSE), the International Conference on Software Maintenance and Evolution (ICSME), and the Empirical Software Engineering journal (Empir. Softw. Eng.) with 3 papers each.
Publication years range from 2017 to 2025 with 4 to 5 papers each year except for 2017 and 2019 (1 paper each) and 2025 with 2 papers at the time of data collection.

The 18 scientific computing papers (SC) include 14 journal papers, 3 conference papers, and 1 textbook, covering a wide range of research areas, including bioinformatics and life sciences (5 papers), computer science (4 papers), psychology (3 papers), and astronomy (1 paper). The remaining 5 papers are general discussions on reproducibility in scientific computing. The publication years range from 2015 to 2024, with 2 to 3 papers each year except for 2015, 2018, and 2024 (only 1 paper each) and 2020 (no paper). The oldest paper in our collection (SC3) has been cited more than 1000 times.

The 7 other papers (O) include 3 practitioners books, 1 magazine article, 1 conference paper, 1 master thesis, and 1 web page.

The results of our content analysis is a taxonomy of beliefs and recommendations that we detail graphically in \Cref{fig:beliefs-SLR} and \Cref{fig:recommendations-SLR}. In the following, we summarize the findings that are most relevant to our research questions. The underline in the text indicates paraphrasing of the focused codes from our taxonomy. We often use direct quotes from the papers to illustrate our focused codes, but note that focused codes are always derived from multiple excerpts across several papers.

\subsubsection{RQ1.a: Docker reproducibility expectations}


Among the main SLR results, we find that the scientific literature contains a wide range of beliefs and expectations about Docker reproducibility, many of which \emph{contradict each other}.
On the one side, \ul{Docker is often framed as a tool for reproducibility}, with bold claims such as ``containerization with Docker \textbf{guarantees} full computational reproducibility'' (SC1, emphasis ours) or ``containers cannot be matched when it comes to enabling reproducibility in a lightweight and portable manner'' (SC8), in particular in the context of scientific computing, where \ul{reproducing the exact environment used to produce results is viewed as essential}: ``full computational reproducibility is only achieved if the software versions used originally are precisely documented.'' (SC1).
These claims are however rather vague in terms of \emph{what} exactly enables reproducibility (is it the distribution of Docker images? of Dockerfiles?).

Other quotes from these and other papers however clarify the \ul{importance of the Dockerfile for reproducibility and transparency}: ``it is [\ldots] essential that the process of creating and building containers themselves is transparent and reproducible'' (SC13).
Yet, the expectations are sometimes that \ul{reproduction of a Docker image from a Dockerfile is straightforward}: ``These containers can be replicated from a single blueprint specified by a text file known as a Dockerfile.'' (SE12) and that \ul{Dockerfiles are precise and unambiguous}: ``There is little possibility of the kind of holes or imprecision in such a script that so frequently cause difficulty in manually implemented documentation of dependencies'' (SC3);
``Dockerfiles are declarative definitions of an environment that aim to enable reproducible builds of the container.'' (SE4).
This is however debated by other authors: ``Dockerfiles are often not easy to follow, and in many cases, do not clarify which specific packages are being deployed'' (SE15).

Sometimes, \ul{sharing a Docker image or Dockerfile is presented as equivalent}: ``There are two ways to share a Docker image; either by sharing the Dockerfile that creates the image or by sharing the image itself [\ldots]. While both ways \textbf{guarantee} a replicable computational environment, sharing the Dockerfile is more transparent and more space-saving.'' (SC1); ``By using Docker containers, developers need only share a Docker image or Dockerfile, which \textbf{guarantees} the environment will be the same'' (SC18).

On the other hand, this is contrasted by quotes that highlight the limitations of Docker for \emph{long-term} reproducibility, which would mean staying reproducible for at least a decade (SC9).
While it is claimed that: ``One of the core promises of Docker is that images are stable environments that can be downloaded and run even years after their creation.'' (SE3), this is also contrasted by the \ul{lack of guarantees regarding long-term reproducibility of Docker images}:
``Ideally, containers that we built years ago should rebuild seamlessly, but this is not necessarily the case'' (SC16).
Even when the Dockerfile still builds, it \ul{may not produce an identical environment}:
``Dockerfile does not guarantee the same build every time'' (SC8);
 ``it is highly unlikely that over time you will be able to recreate [the image] precisely from the accompanying Dockerfile'' (SC16).

Another reason why sharing a Dockerfile is considered important is that it \ul{allows to reproduce the build environment with controlled variation}, which can be important for both auditing and building upon past scientific experiments:
``As some of these changes may indeed resolve valid bugs or earlier problems [\ldots], it will often be insufficient to demonstrate that results can be reproduced when using the original versions'' (SC3);
``The layer of abstraction that a Dockerfile provides makes it easier to change [\ldots] versions or experiment parameters later on.'' (SC2).

\begin{figure}
  \includegraphics[width=\textheight, angle=90]{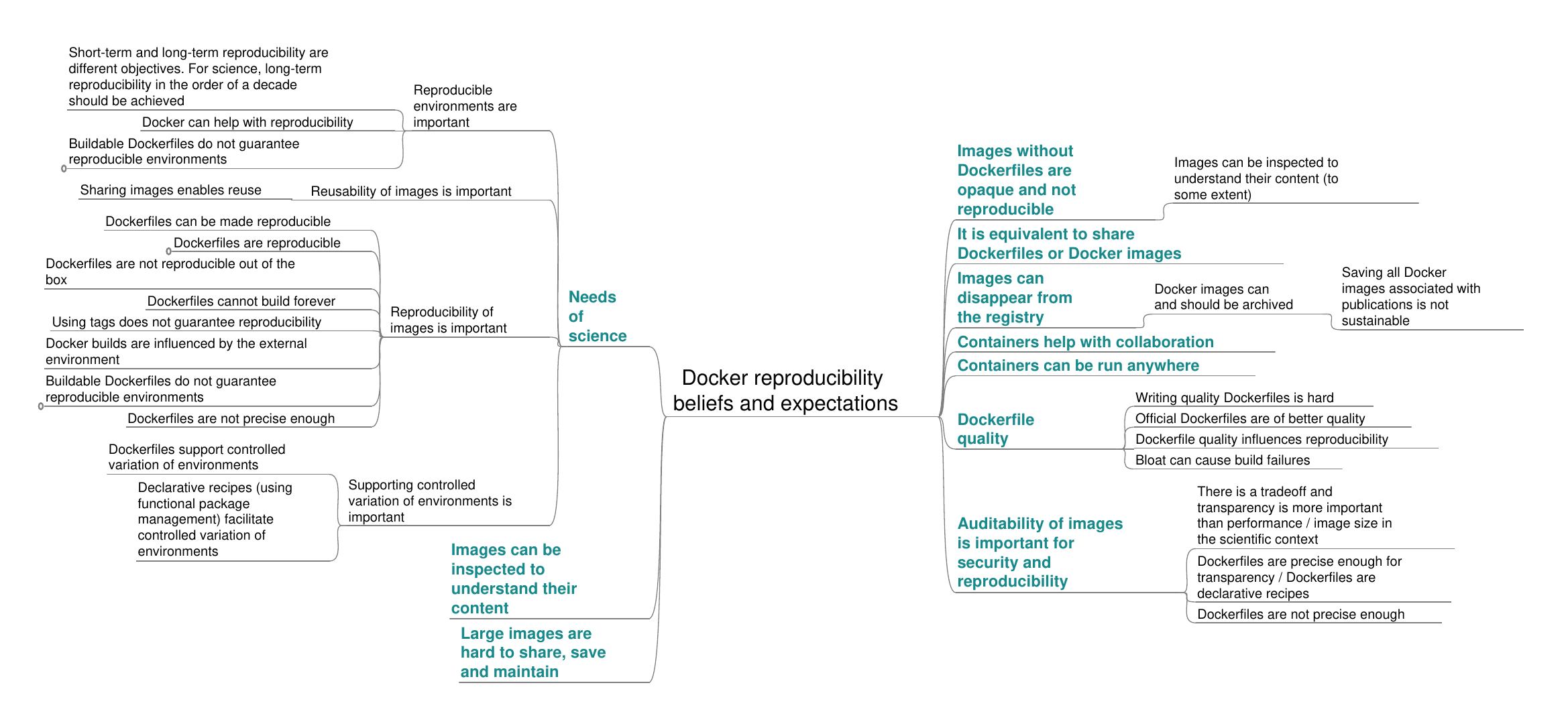}
  \caption{Beliefs and expectations on Docker reproducibility extracted from the literature review.}\label{fig:beliefs-SLR}
\end{figure}

\Cref{fig:beliefs-SLR} summarizes the taxonomy of the diverse beliefs and expectations surrounding Docker reproducibility extracted from our SLR.

\subsubsection{RQ1.b: Causes of non-reproducibility}

The most commonly documented cause of non-reproducibility (which may manifest in terms of a Dockerfile that does not build, or builds but produces a different image) is related to \ul{changes in the external environment}:
``temporal failures that emerge over time due to dynamic changes in external dependencies, such as updates to base images, third-party libraries, or environmental settings---occurring without any modiﬁcations to the Dockerﬁle itself'' (SE13).
In particular, it can be due to \ul{changes in the dependencies}:
``the absence of a concrete version can lead to the usage of a version which is not compatible with the other components of the container, thus to failed builds, or failures hitting surface only at container execution'' (SE4).
\ul{Dependencies, such as the base image, can also disappear}:
``images that are not pulled by anyone from Docker Hub for extended periods of time get purged, and Dockerﬁles are not guaranteed to build indeﬁnitely'' (SC8).
Finally, the Dockerfile in itself \ul{may not be sufficient to reproduce the build, if other important aspects such as the 
build context are not correctly preserved}:
``many of them are not reproducible for builds as [\ldots] their build context ﬁles are missing'' (SE12).

\subsubsection{RQ1.c: Recommendations for Dockerfile reproducibility}

Most of the recommendations discussed in the literature (particularly in software engineering papers) are not directly related to reproducibility, but rather to image size, build time, and security.
Some of these recommendations (such as \ul{avoiding bloat} by not installing unnecessary dependencies) can however impact reproducibility.
More generally, \ul{using linters} to detect Dockerfile smells~\cite{lin_large-scale_2020} is a common recommendation, supported by the belief that ``\ul{Dockerfiles [\ldots] quality may affect the reproducibility} [\ldots] of the resulting image.'' (SE1).
The most commonly used tool in many publications, but also for practitioners, is \texttt{hadolint}~\cite{hadolint-web}: ``Hadolint is currently the reference tool'' (SE24)
and \ul{official images are often presented as a reference for quality}: ``The official status of a repository can be seen as a quality label'' (SE3).

Given the main cause for non-reproducibility described above, the most common recommendation to improve reproducibility is to \ul{pin dependencies} to specific versions, whether they are installed packages, base images, or dependencies pulled from Git repositories: ``If a copy is done via git clone or equivalent, a specific commit or a tagged git version should be specified, never a branch only.'' (SC13).
Yet, there is a \ul{trade-off between pinning dependencies and the ability to get security updates}:
``pinning package versions requires continuous maintenance to keep versions up to date'' (SE24).
This leads to contradictory recommendations in the literature: e.g., for the base image, some recommend pinning by tag or by digest, while others recommend the use of mutable tags (latest, or major versions tags that still receive updates):
``mutating tags for major releases, which bring regular updates in, may be desirable and in other cases stricter reproducibility is desired'' (SE25).
Similarly, some recommend running \texttt{apt-get update} before installing packages, while others recommend \emph{against} it.

This is why the choice of which best practices to follow depends on the objectives, and practices can vary widely: ``These practices might be somewhat different from the practices of generic software developers in that researchers often need to focus on transparency and understandability rather than performance considerations.'' (SC16).
Nevertheless, \ul{images are also viewed as opaque and difficult to audit}, so, for software supply chain security, pinning dependencies can also be seen as a good practice (O1), which can then be balanced by \ul{using tools} to update pinned dependencies regularly.

Among other contradictory recommendations, we found that some authors recommend \ul{reusing base images} to avoid duplicating work, when others \ul{avoid non-official base images}, and prefer reusing and adapting their Dockerfiles instead.
A third route is reusing base images, but after ensuring that they are rebuildable from their Dockerfile, and saving the Dockerfile in the repository.

Best practices for reproducibility also cover including in version control all the files copied in the image, avoiding the use of build arguments, which can make the build process less transparent, documenting the build process, and avoiding to build parts of the image outside the Docker build process: ``It should be noted that building the software outside containers and only including the final binary in the container is considered a bad practice.'' (SE25).

Finally, some authors recommend \ul{using other recipes than a Dockerfile} and associated tools, be it Maven for Java images (O5, but not specifically for reproducibility purposes) or functional package managers, such as Nix or Guix (SC7, SC17, O3).

\begin{figure}
  \includegraphics[width=\textheight, angle=90]{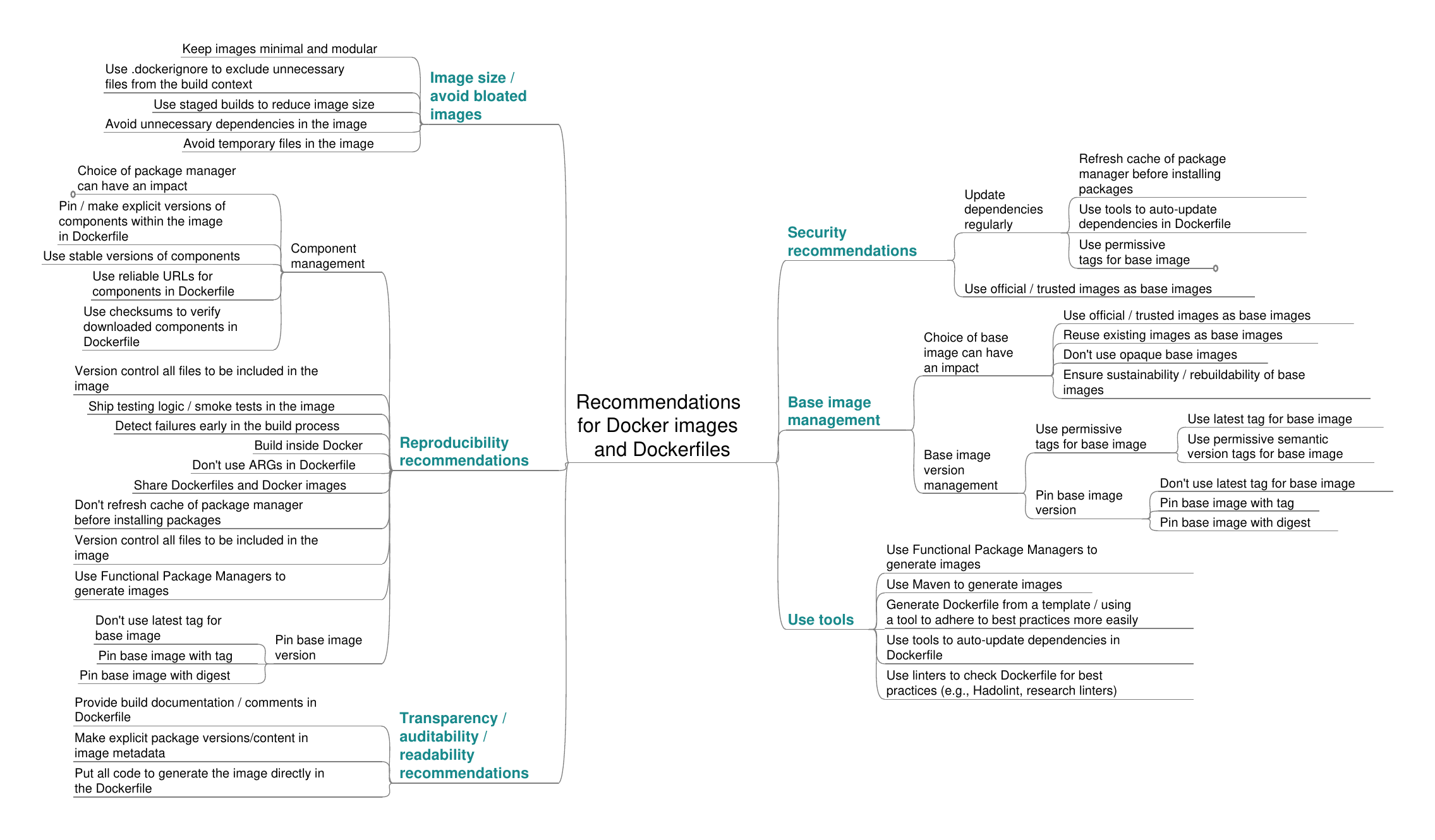}
  \caption{Recommendations on Dockerfiles and images extracted from the literature review.}\label{fig:recommendations-SLR}
\end{figure}

\Cref{fig:recommendations-SLR} summarizes the recommendations on Dockerfiles and Docker images resulting from the SLR.

\section{Experimental Docker image reproducibility}
\label{sec:empirical}

In this section, we describe our methodology for studying Docker image reproducibility empirically, and answer RQ2 and RQ3.

\subsection{Methodology}

\subsubsection{Creation of the dataset}

In order to study the reproducibility of image-building from Dockerfiles, one needs access to historically built Docker images as well as corresponding Dockerfile, source code, and build parameters used to generate them---serving as a ground truth for reproducibility analysis.
While several existing datasets contain Dockerfiles, Docker images, or both, to the best of our knowledge, none provides the precise association between images and the complete build configuration (including repository commit) that produced them.

To address this gap, we leverage the \emph{GHALogs} dataset~\cite{moriconi_ghalogs_2025}, which captures over \num{500000} GitHub Actions workflow runs, along with logs and metadata, collected across \num{25000} software repositories in October 2023. We focus specifically on workflows that use the \texttt{docker/build-push-action}---the official GitHub Action for building and publishing Docker images to repositories. This action's metadata contain all relevant build parameters (e.g., Dockerfile path, build context, arguments, and target stage) necessary to locally reproduce the image build. Besides, when images are pushed to a repository, we can analyze traces left in the execution logs.

\begin{figure}[!htbp]
\centering
\begin{tikzpicture}[
  font=\small,
  node distance=5mm,
  >=Latex,
  box/.style={
    draw,
    rounded corners=2pt,
    align=center,
    inner sep=6pt,
    text width=0.84\linewidth
  },
  final/.style={
    draw,
    rounded corners=2pt,
    align=center,
    font=\small\bfseries,
    inner sep=7pt
  },
  arrow/.style={->, line width=0.6pt}
]

\newcommand{\step}[2]{%
  #1\\[-1pt]
  {\footnotesize\textit{Remaining:} \textbf{#2}}%
}

\node[box] (s0) {%
\textbf{20\,141 candidate runs} containing the \texttt{docker/build-push} action.
};

\node[box, below=of s0] (s1) {%
\step{\textbf{Filter out} 7 runs which build Docker context outside of the repository.}{20\,134}
};

\node[box, below=of s1] (s2) {%
\step{\textbf{Filter out} 10\,298 runs that contain non-reproducible steps \emph{before} the \texttt{build-push} action.}{9\,836}
};

\node[box, below=of s2] (s3) {%
\step{\textbf{Filter out} 2\,993 runs whose \texttt{actions/checkout} action references external repositories.}{6\,843}
};

\node[box, below=of s3] (s4) {%
\step{\textbf{Filter out} 157 runs building images for targets that do not include \texttt{x86\_64}.}{6\,686}
};

\node[box, below=of s4] (s5) {%
\step{\textbf{Filter out} 64 runs that use secrets.}{6\,622}
};

\node[final, below=of s5] (end) {Final dataset:\\ \textbf{6\,622 workflows}};

\draw[arrow] (s0) -- (s1);
\draw[arrow] (s1) -- (s2);
\draw[arrow] (s2) -- (s3);
\draw[arrow] (s3) -- (s4);
\draw[arrow] (s4) -- (s5);
\draw[arrow] (s5) -- (end);

\end{tikzpicture}
\caption{Filtering steps used to construct the workflow runs dataset used in our experiment.}
\label{fig:dataset-filtering}
\end{figure}
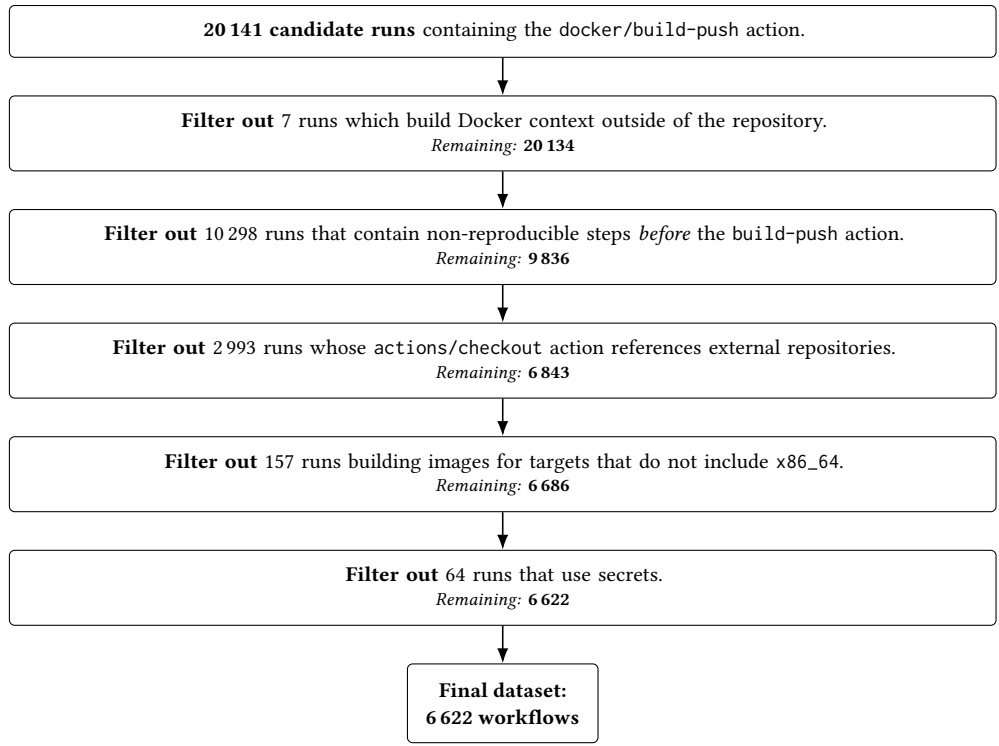

To create our dataset, we apply a series of filtering and cleaning steps to the raw GHALogs data (summarized in \Cref{fig:dataset-filtering}). We first discard all workflow runs that do not invoke the Docker \texttt{build-push} action. From the remaining \num{20141} runs, we examine both the parameters passed to the \texttt{build-push} action and the actions executed prior to it, as preceding actions may affect the environment in which the Docker image is built in ways difficult to reproduce locally and thus impact the result. We list all actions appearing before the \texttt{build-push} step in the remaining runs and rank them by frequency. To assess their potential impact on the Docker build and whether their effects can be reproduced outside GitHub Actions, we review the documentation and source code of the 10 most frequent actions. We exclude actions that set up or modify the build environment outside Docker, focusing on images built entirely within Docker, which we identified as a good practice in our SLR. We retain only actions whose effects we can confidently reproduce locally or that have no effect on the Docker build, namely \texttt{actions/checkout}, Docker’s \texttt{login-action}, \texttt{metadata-action}, \texttt{setup-buildx-action}, and \texttt{setup-qemu-action} (see \Cref{table:step-filtering} for details). We discard workflow runs that include any other action before the \texttt{build-push} step, removing \num{10298} runs (including \num{7643} containing an action that we have manually reviewed and rejected). We additionally filter out some runs based on the parameters of the \texttt{build-push} action, such as runs that do not target the \texttt{x86\_64} architecture, or runs that make use of secrets. Runs that checkout code external to their own repository are likewise filtered. After these filtering steps, we obtain a cleaned dataset of \num{6622} relevant workflow runs, of which \SuccessfulRuns{}~completed successfully between July and October 2023.

\begin{table}[!htbp]
\centering
\caption{10 most frequent GitHub Actions present before the \texttt{build-push} step and filtering rationale.}
\label{tab:top-actions-before-build-push}
\rowcolors{2}{white}{gray!5}
\begin{tabular}{l r c p{5.2cm}}
\toprule
\textbf{Action} & \textbf{\#Runs} & \textbf{Kept} & \textbf{Reason for keeping/rejecting} \\
\midrule

\rowcolor{green!15}
\texttt{actions/checkout} & 17\,177 & Yes &
Checks out repository source code. \\

\rowcolor{green!15}
\texttt{docker/setup-buildx-action} & 10\,295 & Yes &
Configures Docker Buildx; reproducible outside GitHub Actions. \\

\rowcolor{green!15}
\texttt{docker/login-action} & 8\,714 & Yes &
Authenticates to container registries; does not affect the build context itself. \\

\rowcolor{green!15}
\texttt{docker/setup-qemu-action} & 6\,622 & Yes &
Enables multi-architecture builds via QEMU; does not impact the build when building CPU native architecture. \\

\rowcolor{red!15}
\texttt{Shell command} & 5\,770 & No &
Arbitrary shell command, not always reproducible locally. \\

\rowcolor{green!15}
\texttt{docker/metadata-action} & 4\,739 & Yes &
Generates image metadata and tags; does not modify the Docker build context. \\

\rowcolor{red!15}
\texttt{actions/cache} & 837 & No &
Caches files across workflow runs; may introduce non reproducible build inputs. \\

\rowcolor{red!15}
\texttt{actions/setup-go} & 388 & No &
Modifies the environment before the \texttt{build-push} action. \\

\rowcolor{red!15}
\texttt{actions/download-artifact} & 337 & No &
Downloads external artifacts into the workspace; may alter the Docker build context. \\

\rowcolor{red!15}
\texttt{actions/setup-node} & 311 & No &
Modifies the environment before the \texttt{build-push} action. \\

\bottomrule
\end{tabular}
\label{table:step-filtering}
\end{table}

\paragraph*{Collection of historical images}

To form our ground truth, we parse the build logs included in GHALogs, extracting both the destination repository where each image was pushed and its \texttt{SHA256} digest, which uniquely identifies the image content.
This approach allows us to retrieve the original push location of \dataDockerImages historical images.
The remaining cases mostly correspond to workflows that built images without pushing them to a registry.
Among these \dataDockerImages images, we successfully retrieved 1541; the remaining 2079 were no longer accessible at the time of our analysis. This number is not unexpected, as images are routinely deleted from registries.

\subsubsection{Empirical study}

We locally rebuild all images that historically built successfully.
All relevant parameters from the \texttt{build-push} and \texttt{setup-buildx} actions such as build arguments, context, target, are passed to the \texttt{docker build} command.
However, variations in time, Docker versions, host operating system, and finally hardware are inevitably introduced, which we consider to be acceptable when testing for reproducibility.  While historical builds have been performed by GitHub runners and in most part on hardware provided by GitHub, we use our own pipeline, on our own \texttt{x86\_64-linux} hardware, to facilitate customization and analysis. Each image is built in isolation within a fresh virtual machine to prevent cross-contamination between builds and to eliminate interference from the host environment.
We use the return code of the \texttt{docker build} command to classify the build as a success or a failure, and export the resulting image in case of success. In any case, we keep the complete build log to identify transient failures (like disk space exhaustion) and re-run them, and to allow future analysis.

\paragraph*{Bitwise reproducibility analysis}

We perform a file-by-file comparison of each historical image against its rebuilt version.
For each pair of historical image matched with the rebuilt one, we walk through the image file tree and compare the content of each file. We count the proportion (over the number of files present in either image of the pair) of files with different content, or files that are missing from one of the images. 

\paragraph*{Functional reproducibility analysis}

To assess whether rebuilt images provide the same software environment as their historical counterpart, we analyze the packages installed in the images and their versions. We use Trivy~\cite{trivy-web}, an open source tool that inspects container images and extracts package metadata from package managers to produce a software bill of material. Trivy supports a wide range of system and ecosystem-specific package managers, allowing us to achieve a good coverage of analyzed images. Still, some projects strip all package management metadata from their container images to minimize bloat which prevents the analysis. We run Trivy on both the historical and rebuilt images, when both exists, setting a timeout of 25 minutes. We collect the extracted package lists for comparison. This step works for \TrivyWorks{}~of the \TrivyTotal{}~images for which both versions were available. We refer to the \TrivyNotAuditable{}~images where Trivy is not able to extract package lists, or a timeout is reached before Trivy completes as ``non auditable''. Using the extracted information, for each pair of images, we compute four metrics, comparing image content at the package level. From the strictest, to the most permissive, we compute the proportion (over the number of packages present in either image of the pair) of packages:
\begin{enumerate}
\item with identical versions;
\item with different versions but identical major and minor versions;
\item with different minor versions but identical major versions;
\item with different major versions but still present in both images.
\end{enumerate}

When checking for identical major or minor versions, we split version numbers by components and keep only the first or the first two components.

\paragraph*{Reproducibility rules detection}

We examine the Dockerfiles in our dataset for reproducibility related issues and best practice violations identified by the SLR. Following common recommendations, we use \texttt{hadolint} as the primary tool to detect Dockerfile quality issues that may impact reproducibility. To select the relevant \texttt{hadolint} rules, the first and last authors reviewed the complete set of available rules and, based on the outcomes of the SLR, performed a consensus-based selection of rules that correspond to one or more recommendations identified in the SLR. Importantly, this selection did not involve a judgment of the intrinsic relevance of individual rules for reproducibility, but merely their correspondence to the recommendations extracted from our SLR. We also implement additional rules corresponding to recommendations not directly covered by \texttt{hadolint} and easily identifiable at the Dockerfile or context level. The resulting set of rules is summarized in \Cref{table:repro-rules}.

\begin{table}
\caption{Reproducibility recommendations extracted from the literature and tested against our Dockerfiles using Hadolint rules (DL) or custom rules (CR).}
\label{table:repro-rules}
\begin{tabularx}{0.72\textwidth}{l X}

\textbf{Rule code} & \textbf{Recommendation that triggers the rule \mbox{violation} when not respected} \\ \hline\hline

\multicolumn{2}{l}{\emph{Pin Dependencies}} \\ \hline
DL3006 & Tag images versions explicitly. \\
DL3007 & Do not use \texttt{latest}. \\
DL3008 & Pin versions in \texttt{apt-get install}. \\
DL3013 & Pin versions in \texttt{pip}. \\
DL3016 & Pin versions in \texttt{npm}. \\
DL3018 & Pin versions in \texttt{apk add}. \\
DL3028 & Pin versions in \texttt{gem install}. \\
DL3033 & Pin versions in \texttt{yum install}. \\
DL3035 & Do not use \texttt{zypper dist-upgrade}. \\
DL3037 & Pin versions in \texttt{zypper install}. \\
DL3041 & Pin versions in \texttt{dnf install}. \\

\hline
\multicolumn{2}{l}{\emph{Avoid Temporary Files}} \\ \hline
DL3009 & Delete the \texttt{apt-get} lists after installing. \\
DL3010 & Use \texttt{ADD} for extracting archives into an image. \\
DL3032 & Run \texttt{yum clean all} after installing. \\
DL3036 & Run \texttt{zypper clean} after installing. \\
DL3040 & Run \texttt{dnf clean} after installing. \\
DL3042 & Avoid cache directory with \texttt{pip install --no-cache-dir} \\
DL3060 & Run \texttt{yarn cache clean} after installing. \\
  CR01 & Use multi-staged builds. \\

\hline
\multicolumn{2}{l}{\emph{Avoid Unneeded Dependencies}} \\ \hline
DL3015 & Avoid additional packages by specifying \texttt{--no-install-recommends}. \\

\hline
\multicolumn{2}{l}{\emph{Fail Early}} \\ \hline
  DL4006 & Set the \texttt{SHELL} option \texttt{-o pipefail} before \texttt{RUN} with a pipe in it \\
\hline
\multicolumn{2}{l}{\emph{Use official images}} \\ \hline
  CR02 & Use images from the official docker library \\

\hline
\end{tabularx}
\end{table}

\paragraph*{Statistical analysis}

We perform a statistical analysis to understand whether there is a statistical dependency between the Dockerfile quality recommendations identified in the literature and our measured reproducibility outcomes for the corresponding Docker images. Since the recommendations may be correlated to one another, testing each one independently for statistical significance would risk overlooking confounding effects. To account for potential correlations, we perform regressions, using the presence or absence of each individual rule as input variables and the reproducibility metric under study as the output variable. This approach allows us to address two questions:
\begin{enumerate}
\item Is there a statistically significant relationship between violations of reproducibility-related recommendations and the observed reproducibility of Docker images?
\item If so, which individual recommendations are significantly associated with more reproducible outcomes?
\end{enumerate}

\paragraph*{Comparison with projects from the official Docker library}

SLR results highlighted two common beliefs in academic discourse: (1)~Dockerfile quality impacts reproducibility, and (2) Dockerfiles of official images are authored by Docker experts, and thus are of particularly high quality. To evaluate these claims, we extracted Dockerfiles from the \texttt{docker-library/official-images} repository, which tracks all versions of the Dockerfiles used to build official library images. In order to have a comparable time drift between the Dockerfiles of official images and the rest of the Dockerfiles used in our empirical study, we took a snapshot of this repository from October 2023, matching the timeframe of the collection of the rest of our Dockerfiles. We then rebuild the collected Dockerfiles using the same experimental pipeline as the community-sourced ones, allowing us to directly compare the reproducibility of expert-authored Dockerfiles with those mined from the wider ecosystem.

\subsection{Results}

We present our experimental results below, organized by research question.

\subsubsection{RQ2: Levels of reproducibility of built images}

\begin{figure}[t]
  \centering

  \begin{subfigure}{0.8\textwidth}
    \centering
    \includegraphics[width=\linewidth]{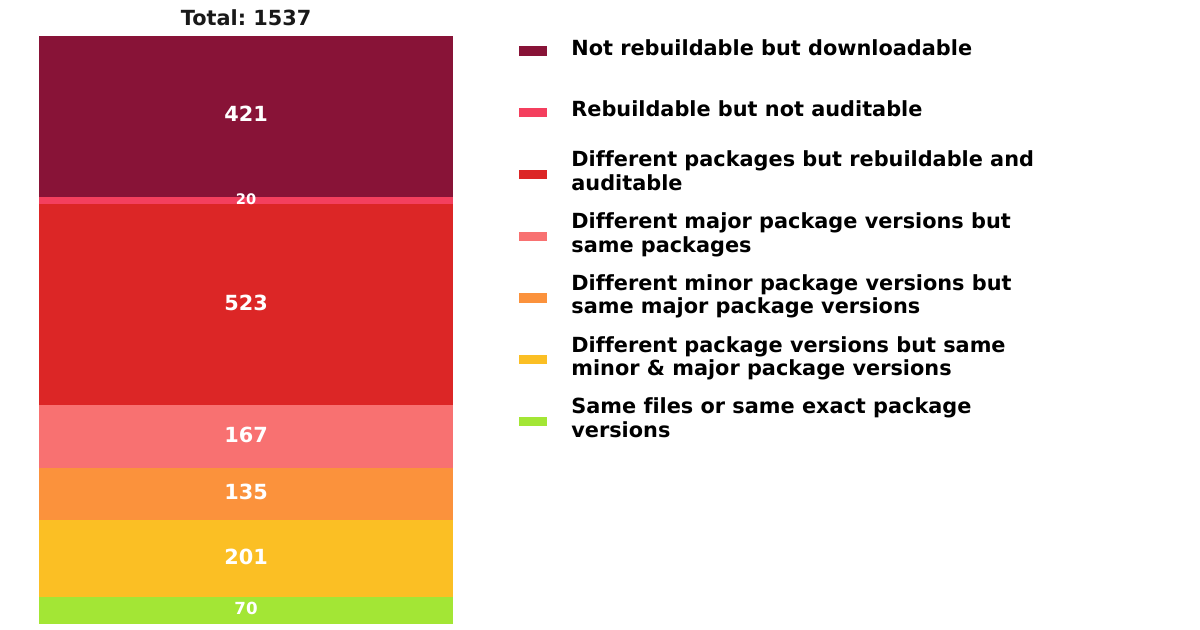}
    \caption{Stacked representation. The “same file” metric (only \ImagesBitwiseReproducible{}~reproducible images) is merged into the “same packages” metric (\ImagesFuncExactReproducible{}~reproducible images) for readability.}
    \label{fig:metrics-stacked}
  \end{subfigure}

  \vspace{1em}

  \begin{subfigure}{0.8\textwidth}
    \centering
    \includegraphics[width=\linewidth]{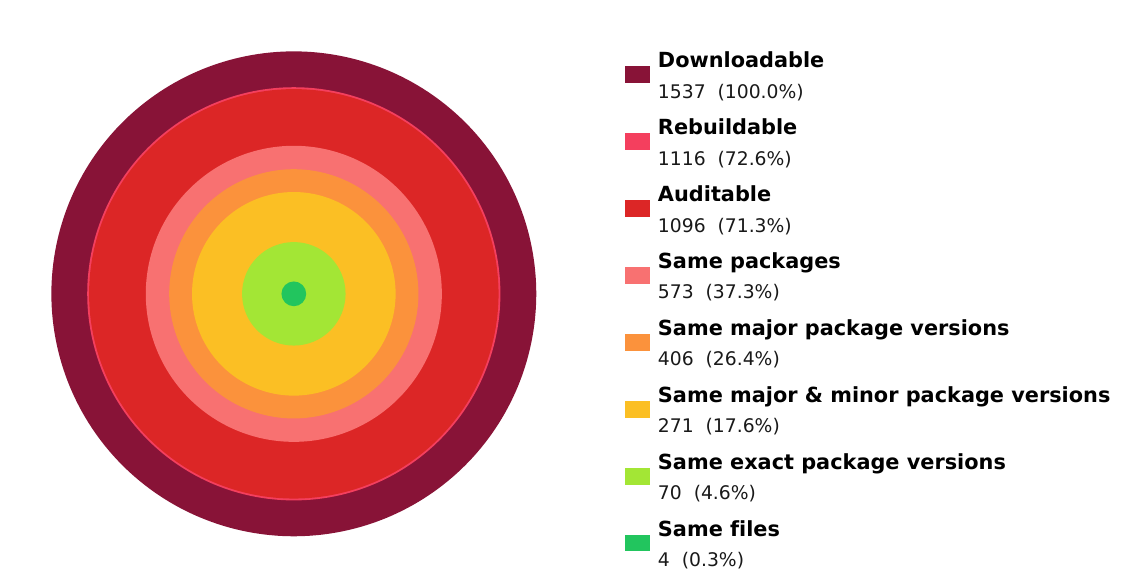}
    \caption{Euler diagram representation.}
    \label{fig:metrics-nested-euler}
  \end{subfigure}
    \caption{Number of images fully verifying each reproducibility metric, among the images for which we could retrieve the historically built version.}

  \label{fig:metrics-combined}
\end{figure}

\begin{figure}
  \includegraphics[width=0.8\textwidth]{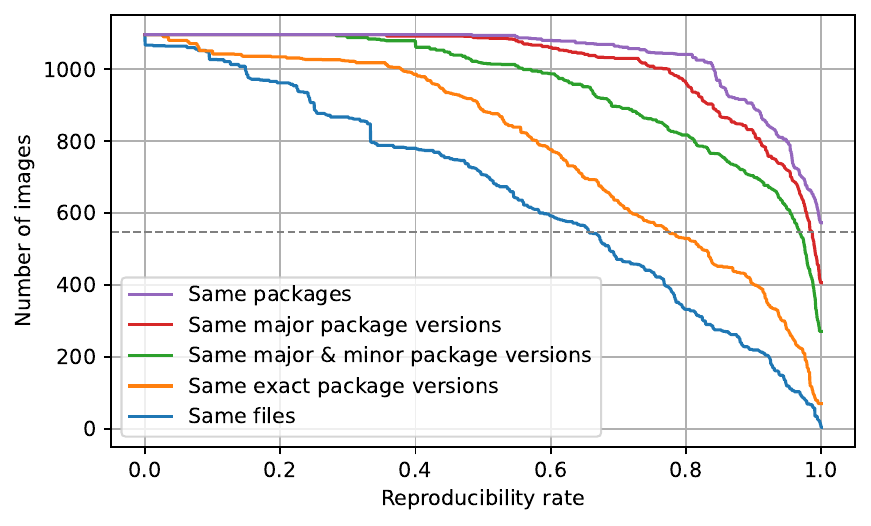}
  \caption{Number of images achieving at least a given reproducibility rate on each of our four metrics.}\label{fig:metrics-cumul}
\end{figure}

We report on the level of reproducibility measured on the Docker images in our dataset. We consider an image to be fully reproducible under a certain metric when all \emph{packages} (resp.~\emph{files} for bitwise reproducibility) are reproducible. \Cref{fig:metrics-combined} summarizes the number of images that are fully reproducible under each metric. Additionally, for the images that rebuilt successfully, we discuss the proportion of packages reproducible for each metric, starting with the strictest definition of reproducibility and progressively relaxing it towards more functional metrics.
\Cref{fig:metrics-cumul} shows the number of images achieving at least a given reproducibility rate for each metric.

\paragraph*{Bitwise reproducibility}

Although the default Docker builder supports the \texttt{SOURCE\_DATE\_EPOCH} environment variable since February 2023 to set the date metadata of the image layers, none of the workflows in our dataset use it, and as a consequence none of the images we built have the same output hash as their historical counterpart. When looking at files content only, we find that only \ImagesBitwiseReproducible{}~images have all their files bitwise reproducible with their historical version, and half of the images that were successfully rebuilt in our dataset have more than \MedianBitwiseComp{}\% of files with differing hashes.

\paragraph*{Functional reproducibility}

When looking at images through the package lens, we still observe very low reproducibility rate with at least~\MedianFuncExactComp{}\% of packages changing version between the original and rebuilt images in half of the cases. Comparing package versions this way, only~\ImagesFuncExactReproducible{} of the compared images achieve full reproducibility.

Although when comparing minor versions, we get a drastically lower proportion of non-reproducible packages---half the images have less than \MedianFuncMinorComp{}\% differing packages---still only a minority of~\ImagesFuncMinorReproducible{} out of~\TotalImagesFuncComp{} images are reproducible under this prism.
This number increases slightly when considering only major version changes, where~\ImagesFuncMajorReproducible{} images are reproducible, with a median of \MedianFuncMajorComp{}\% packages changing major version. Finally, \ImagesFuncPackagesReproducible{} images out of~\TotalImagesFuncComp{} contain the same packages as their counterpart, without accounting for versions.

\paragraph*{Rebuildability}

Finally, the most minimal expectation for reproducibility is being able to rebuild an image at all. While all~\SuccessfulRuns{}~jobs that we tried to build did produce an image in July--October 2023, we find that less than two years later only~\NumberOfImagesBuilding{} are still building successfully, hence resulting in a~\PropOfImagesBuilding{}\% rebuildability rate. Importantly, an inspection of the build logs suggests that failures due to missing base images are relatively rare: only 43 failed jobs can be attributed to an unavailable base image, including just 2 cases that may be explained by the use of a private base image. This observation mitigates concerns that a substantial fraction of the selected runs would be unreproducible simply because they depend on private base images.

\subsubsection{RQ3: Effectiveness of recommendations for reproducibility}

\begin{figure}
  \includegraphics[width=0.8\textwidth]{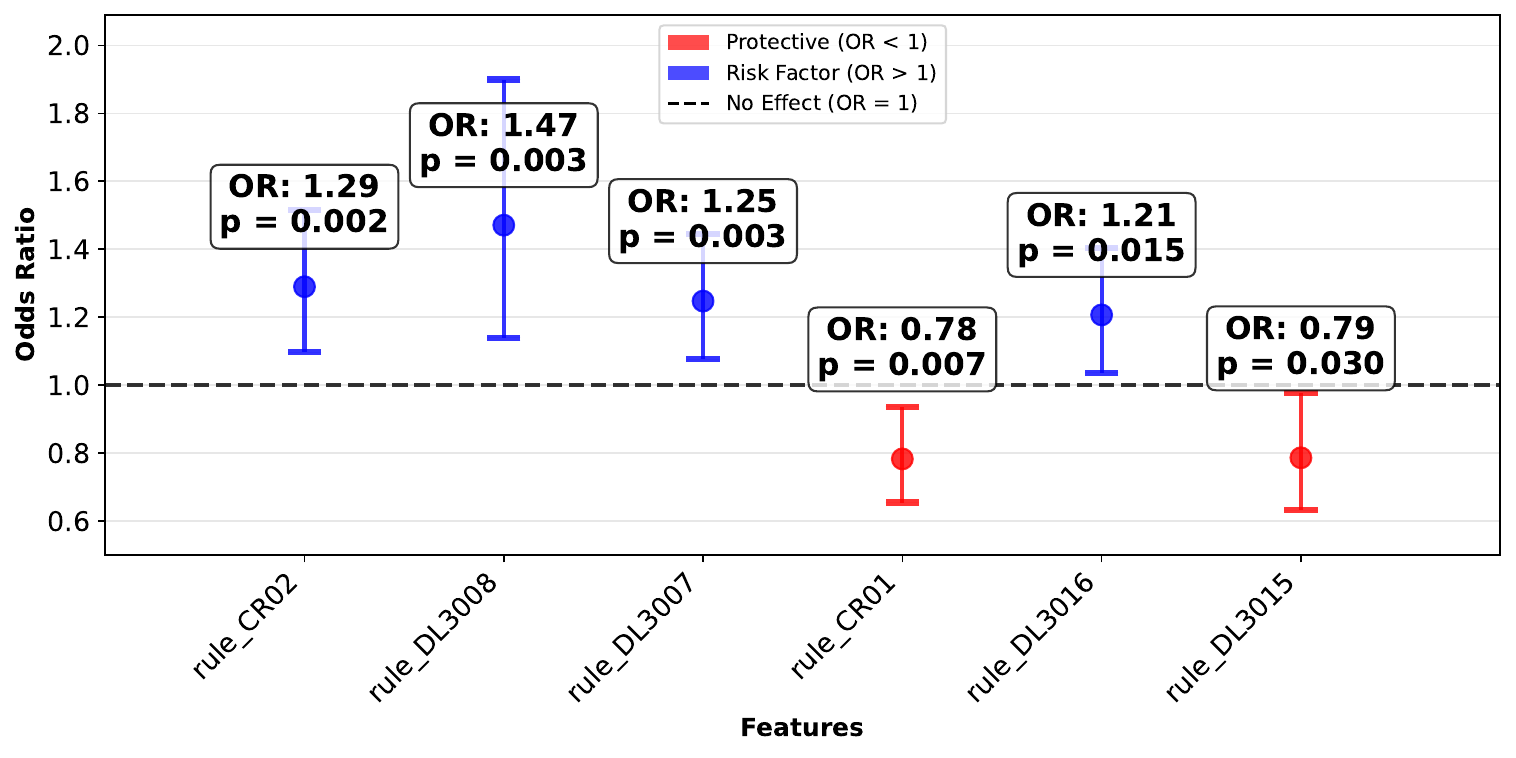}
  \caption{Significant rules for rebuildability: blue variables are rules that decrease rebuildability when violated, red ones that increase it. Rules \texttt{DL3008}, \texttt{DL3007} and \texttt{DL3016} are pinning rules.}
  \label{fig:ci-rebuild}
\end{figure}

We perform regression analyses to understand whether there is a statistically significant relationship between the reproducibility recommendations we extracted from the literature and the various reproducibility outcomes observed in our study. For each regression, the input variables are binary indicators representing for each rule whether it is violated. The output variable is either the proportion of unreproducible packages---used in linear regressions for package-level reproducibility---or the build outcome (success or failure), modeled using logistic regression for rebuildability. Before performing the regressions, we filter out rules that appear in less that 1\% of our examples, because they may lead to overfitting.
We observe a p-value lower than $0.05$ for each of our models, showing a statistically significant dependency between the recommendations studied and our metrics. Furthermore, for each metric, some of the rules tested also have a statistically significant effect. However, the fit of our models, with $R^2$ respectively of \RRebuildability{} for rebuildability, \RBitwise{} for file reproducibility, \RFuncExact{} for exact package version reproducibility, \RFuncMinor{} for minor version reproducibility, and \RFuncMajor{} for major version reproducibility shows a poor explanation power of the models on variability of the reproducibility rates.  Furthermore, interpretation of the individual rules coefficients is also hazardous as, for all but the rebuildability metrics, the rules fitted have very weak effects, explaining only a few percentage points of the variability. In that context, we even see in some cases counter-intuitive results with some pinning rules appearing to contribute negatively to the reproducibility rate for some metrics, but with very small effect size (below 3\%). This suggests that these models are only measuring a weak signal and that there might be confounding factors at play, like the evolution of the base image used or of the system package manager. We show in \Cref{fig:ci-rebuild} the statistically significant rules for our rebuildability model, and all other regressions results are available in our replication package~\cite{this-replication-package}.

\paragraph*{Reproducibility of the Docker official images}

Another angle to understand if quality differences in Dockerfiles may have an impact on reproducibility is to test Dockerfiles expected to be of high quality. That is what we do by rebuilding the 121 Dockerfiles from 2023 extracted from Docker official images, successfully rebuilding 106, that is 88\%. This 16 percentage points increase in rebuildability rate compared to the Dockerfiles extracted from the wider ecosystem is consistent with the beliefs that we identified in our SLR that the official Dockerfiles are of higher quality and that the quality of the Dockerfile correlates with better rebuildability. 

\section{Discussion}
\label{sec:discussion}

\subsection{Takeaways}

In this work, we have looked at the question of Docker images reproducibility through the prism of six different reproducibility levels tied to varying expectations and needs. Our empirical findings show that \textbf{Docker provides no guarantees for any of these reproducibility levels}.
This can break some assumptions held by the scientific community, as highlighted by our systematic literature review (SLR).
Our results show that sharing a Dockerfile is not equivalent to sharing a built Docker image, and that with time the content of the image may change, increasing the chances of impeding the functional behavior of the image when the software environment changes too much.
Even more symptomatic, although not necessarily worse, images may even stop building altogether when external resources go missing or the build environment varies significantly.

For software engineering purposes, these concerns compound with another one: the inability of Docker to generate bitwise identical images \textbf{forces users to blindly trust the images they are downloading to be faithfully generated} from their documented sources.
In a context where high profile attacks on the software supply chain become more common~\cite{ladisa-2023-ssc-taxonomy}, this lack of proof of correspondence between source code and generated image places Docker at a disadvantage compared to other means of software distribution that provide such guarantees, for example some Linux distributions like Debian or Nixpkgs.
Although we have shown that there is a statistically significant effect of some of the recommendations found in the literature about image reproducibility, it is to be put in perspective with the modest size of the aforementioned effects.
This suggests that \textbf{following best practices to write Dockerfiles alone helps, but is not a silver bullet} for achieving image reproducibility in Docker-based workflows.

Our findings should not be seen as a call to abandon neither containerization technologies nor Docker.
These tools fulfill important roles in software engineering: they provide portability, simplify deployment, and enable execution environment isolation.
However, we believe that users and projects relying on Dockerfiles to build container images should critically assess their Docker usage in light of our results. If image reproducibility is an important concern for their workflows, our findings suggest they may need to adjust their practices---or even reconsider their tooling---to better align with their reproducibility goals. 
Users who cannot easily switch to alternative tools should consider adopting the best practices we experimentally validated as contributing positively to reproducibility, such as pinning dependencies whenever possible. However, they should not do so with the expectation that this will \emph{guarantee} reproducibility of their images.

Our SLR also highlights \textbf{alternative approaches---such as the use of functional package managers}---whose reproducibility guarantees, both for rebuildability and bitwise reproducibility, are backed by recent experimental results~\cite{malka-repro-space-time-2024, malka_does_2025}. Notably, functional package managers remain usable with the Docker runtime, as they can generate OCI-compliant images while preserving the reproducibility benefits of their model.

In reviewing the literature, we were struck by the diversity---and at times, \textbf{contradiction---of the beliefs and recommendations} we encountered, occasionally even within the same article. We also observed that some papers used vague or imprecise terminology, which often led to questionable or unsupported claims. Furthermore, and this justifies our use of the term “beliefs,” we found that few of the assertions promoting Docker as a tool for reproducibility or for use in scientific workflows were actually supported by empirical evidence or grounded in research.
This is not caused by a disconnect between research fields to the point that empirical research on Docker would not be read or taken into account by scientists in other disciplines, but rather by the \textbf{extreme scarcity of software engineering work on this topic}.
This is concerning, as it suggests that technical decisions shaping scientific practice are being made without input from software engineering domain experts.
Such a disconnect carries the risk of contributing to systemic issues like the reproducibility crisis.
It is our opinion that it is the responsibility of software engineering researchers to support the broader scientific community by sharing their expertise and helping guide the adoption of sound technical practices.
It is in this spirit that we undertook this work.

\subsection{Threats to validity}

We adopt the terminology of Runeson et al.~\cite{runeson_guidelines_2009} in this section.

\subsubsection{Construct validity}


A first concern is that the SLR could have missed some relevant papers, in particular by the choice of search keywords. This concern is mitigated by the use of Google Scholar full-text search to cover all types of papers, complemented by the backward snowballing. Furthermore, according to Ralph et al.~\cite{ralph_paving_2022}, this concern is not as relevant for critical reviews, where a subset of papers can be sufficient to highlight problems in the analyzed area. In our case, the papers we analyzed were indeed sufficient to reveal many contradictory statements and unfounded beliefs.

Another concern regarding the SLR is that we may have misinterpreted some of the papers, or that the quotes we extracted do not represent the authors' intent. However, it is important to note that even if these quotes may not perfectly represent the \emph{authors' view}, they carry weight as they can influence what other researchers and practitioners will believe about Docker reproducibility based on what is \emph{written}. Therefore, it is important to be able to highlight when some of these statements are misleading.

Regarding our empirical study, the main risk is that our convenience sample for testing reproducibility from the GHALogs dataset is not representative of the wider population of Dockerfiles and Docker images. Indeed, we cannot claim that the figures that we obtained would be the same on different samples, and we even show this by comparing to a gold set of Dockerfiles from the official Docker library. Nonetheless, the exact reproducibility rates are less important than our main finding: that reproducibility of image building is not guaranteed by Docker out of the box.

\subsubsection{Internal validity}


When relating smells in Dockerfiles to the reproducibility of the images built from them, we control for the possibility that smells are correlated with each other, by performing a regression analysis that takes into account all smells at the same time.
However, it is very likely that some confounding factors that cannot be observed directly are at play.
Therefore, we make no claim of causal effects in our results, and we discourage the reader to misinterpret them as showing that some recommendations are proved to be effective.
Besides, while some recommendations are indeed correlated with the reproducibility of the output in our regression results, the effect sizes are quite small, which means that these recommendations alone are not sufficient to achieve the reproducibility objective.

\subsubsection{External validity}


As already noted, we do not claim that our figures would generalize to other sets of Dockerfiles. Furthermore, even when using the same sample, these results cannot generalize in time. We performed our reproducibility study less than two years after the original builds, and it is likely that we would have obtained higher reproducibility rates if we had performed the study earlier, and that one will obtain lower rates in the future.

\subsubsection{Reliability}


For the SLR, we used constructive grounded theory~\cite{charmaz_constructing_2014} to derive a taxonomy of beliefs and recommendations. This methodology does not adhere to the positivist view of reliability, and therefore other researchers could have derived a different taxonomy. The way to judge this research is therefore not whether it can be repeated yielding identical results, but whether the results are useful and grounded in the data we collected. To allow closer inspection of our research process and results, we provide detailed information about it in the Zotero collection that we used for the SLR~\cite{this-slr-zotero}, and in our replication package~\cite{this-replication-package}.

For the empirical study, our replication package~\cite{this-replication-package} includes all code used for the analysis. It can be used to replicate our study, subject to the time-related caveat described above. However, due to data size, we do not provide all intermediate outputs, such as built images.

\section{Related work}
\label{sec:related}

\subsection{Docker studies}

In our work, we have collected data from Docker builds, Dockerfiles, and Docker images, and we have run an experiment consisting in rebuilding Docker images from their Dockerfiles.
There are many studies in empirical software engineering that focus on Docker, and which have done one or more of these steps.
To the best of our knowledge, however, none of them has done all of them together, and few have focused on reproducibility.

\paragraph{Collecting Dockerfiles}


Many studies have mined Dockerfiles from public repositories (most often from GitHub) to study their quality, usage, and evolution.
Among the earliest work on the topic, Cito et al.~\cite{cito_empirical_2017} collected \num{70000} Dockerfiles from GitHub to study their quality, using \texttt{hadolint} to detect smells.
This is also the case of Henkel et al. who collected \num{178000} Dockerfiles from GitHub~\cite{henkel_dataset_2020}, and later used them as an empirical basis for the design of a new linter (Binnacle)~\cite{henkel_learning_2020}, or Zhou et al.~\cite{zhou_drive_2023} who similarly collected Dockerfiles to derive rules for a new linter (DRIVE).
We initially considered running these two linters in addition to \texttt{hadolint} in our empirical study, but we found that none of them was properly packaged for reuse (no installation or usage instructions).
More recently, Durieux collected Dockerfiles and manually annotated a subset to serve as a ground truth for evaluating linting and repair tools~\cite{durieux_empirical_2024}. His own tool, \texttt{Parfum} is however focused on smells related to image size.
Finally, the largest study of Dockerfiles to date is by Eng et al., where they used World of Code to analyze 9 million Dockerfiles, and study their evolution over time~\cite{eng_revisiting_2021}.

\paragraph{Collecting Docker images}

Other authors have collected Docker images from public registries (generally Docker Hub). For instance, Zerouali et al.~\cite{zerouali_multi-dimensional_2021} collect Docker images from Docker Hub to analyze technical lag.
The notion of technical lag~\cite{gonzalez-barahona_technical_2017} is close to the metrics that we use for evaluating functional reproducibility, as they share the idea of measuring and aggregating a drift in versions of packages present in an image. The main difference is that technical lag is a measure of the distance to an ideal version (usually the latest), while our functional reproducibility metrics are a comparison between two images, one historical and one rebuilt.

Sometimes, Docker images are associated with (GitHub) repositories containing Dockerfiles. That is the case in the study by Lin et al.~\cite{lin_large-scale_2020} that investigates at the same time characteristics of Docker images (such as their size) and Dockerfiles (such as the smells they contain).
None of these datasets, however, provide precise links between a specific version (repository commit) of a Dockerfile and the built image, nor the exact build parameters, which are essential for reproducibility analysis.

\paragraph{Collecting Docker build results}

Some studies collect Docker build data, usually from the Docker Hub automated builds, and not from GitHub workflows like we did.
Wu et al.~\cite{wu_empirical_2020} collected Docker build statuses, but their study focuses on how often build failures occur, how long it takes to fix them, and the evolution of these metrics over time. In later work~\cite{wu_understanding_2022}, they also look at build durations. Similar to us, they correlate characteristics of Dockerfiles (including \texttt{hadolint} smells) with build duration. However, they do not try to rebuild images.
The same can be said of a study by Zhang et al.~\cite{zhang_insight_2018} that also correlates build duration and Dockerfile characteristics.

\paragraph{Rebuilding Docker images}

A few studies before ours have attempted to rebuild Docker images from Dockerfiles. Cito et al.~\cite{cito_empirical_2017} did this for a random sample of 560 Dockerfiles from their dataset mentioned above, and observed a 34\% failure rate. Similar to us, they observed more build successes when Dockerfiles used image pinning. However, in their study, the original build status of the Dockerfile had not been collected. Therefore the failed builds could also have been from Dockerfiles that had never built successfully in the first place, or by not using the right build arguments. Besides, each Dockerfile smell was correlated with build success rate independently of the others, which is not as reliable to draw conclusions, because smells can be correlated with each other.

Shabani et al.~\cite{shabani_dockerfile_2025} also rebuilt Dockerfiles but, in their case, with the goal of studying the flakiness of Docker builds. They started by building a large number of Dockerfiles (about \num{18000}), and observe a 55\% failure rate (with the same limitations as Cito et al.). However, they then proceeded to rebuild the Dockerfiles that built successfully every week for 9 months, and observe the failures. Interestingly, while the aim of the study was to study flakiness, which is normally understood as a transient failure, they report a large proportion of ``permanent flakiness'', i.e., Dockerfiles that used to build and then stop building after some time. Although their focus was not reproducibility, and the authors did not measure the semantic drift of rebuilt images, these results go in the same direction as ours.

Recently, Guillauteau et al.~\cite{guilloteau_longitudinal_2025} introduced an initial study design aimed at assessing the reproducibility of software environments defined by Dockerfiles in research artifacts.
While their work differs from ours in both methodology and the scale of the empirical evaluation, it shares a similar objective: highlighting the limitations of Dockerfiles as a mechanism for long-term reproducibility, particularly in scientific contexts.

Finally, Zhu et al.~\cite{zhu_doctor_2025} also rebuilt Dockerfiles to evaluate the optimization they apply to them, but they do not report on the initial failure rate of the collected Dockerfiles.

\subsection{Reproducibility studies}

The so-called ``reproducibility crisis''~\cite{hassan_characterising_2025, perkel_challenge_2020} in empirical sciences is largely documented and has led to many studies focusing on the reproducibility of scientific experiments. We do not aim to review these works here. Instead, we focus specifically on the software engineering literature on reproducibility.

Reproducibility is an important concern in software engineering for security~\cite{lamb_reproducible_2022}, collaboration workflows, maintenance~\cite{ren_automated_2018}, software preservation~\cite{courtes-guix-swh-2024}, and applications to scientific computing.

In the context of the latter, Hassan et al.~\cite{hassan_characterising_2025} performed a systematic literature review on \emph{reproducibility debt}, which they define as technical debt affecting scientific software and the resulting ability to reproduce scientific results. While they do not specifically focus on the reproducibility of software environments, several of the causes of reproducibility debt that they report are connected to this aspect (e.g., missing dependencies, issues with software versions).

In previous work~\cite{malka-repro-space-time-2024}, we investigated reproducibility of build environments in the context of the Nix functional package manager.
In a later study~\cite{malka_does_2025}, we used an experimental approach similar to the one of this paper, consisting in re-running builds for which a historical truth is still available.
However, the focus of this study was on bitwise reproducibility, which we have shown can be achieved at a large scale using Nix.
In the context of the present study, most Docker images were far from bitwise reproducible, so we had to resort to more relaxed definitions of reproducibility.
Other relaxed definitions have been investigated before,
e.g., by Pöll et al.~\cite{poll_analyzing_2021}, with the concept of \emph{accountable builds}, in which differences between builds that can be explained are tolerated, or by Sharma et al.~\cite{sharma_canonicalization_2025}, who apply a canonicalization step to build artifacts of unreproducible builds.

There are more experimental studies of reproducibility~\cite{benedetti_empirical_2025,randrianaina_options_2024}. However, these studies most often rebuild the same software twice, rather than comparing a build today, with a historical build.

Finally, while containerization raises the question of whether images can be built reproducibly, as we investigated here, it can also be seen as a tool for reproducible builds, as shown by Navarro et al.~\cite{navarro_leija_reproducible_2020} with their \emph{reproducible containers}. Fundamentally, containerization is tied to the concept of sandboxing, which is also what functional package managers use to improve reproducibility.

\section{Conclusion}
\label{sec:conclusion}

In this work, we investigated the reproducibility of Docker images built from Dockerfiles. Through a survey of the literature, we uncovered beliefs expressed in the scientific discourse about Docker images reproducibility---including that Dockerfiles do provide reproducibility---and recommendations associated to image reproducibility.  We empirically tested these claims by rebuilding \SuccessfulRuns historical images from 2023 GitHub actions workflows. Our results show that Docker does not guarantee reproducibility under any tested definition, nor is there a “silver bullet” set of rules for writing Dockerfiles yielding reproducible images.

\clearpage
\bibliographystyle{ACM-Reference-Format}
\bibliography{bibliography}
\end{document}